# MAC Protocols Design for Smart Metering Network

**Yue Yang[1], Yanling Yin[2], Zixia Hu[1]**

[1]Electrical Engineering, University of Washington, Seattle, Washington, USA
[2]Electrical Engineering, Harbin Engineering University, Harbin, USA

**Email address:**
yueyang@uw.edu (Yue Yang), yinyanling@hrbeu.edu.cn (Yanling Yin), huzixia1984@gmail.com (Zixia Hu)

**Abstract:** The new generation of power metering system - i.e. Advanced Metering Infrastructure (AMI) - is expected to enable remote reading, control, demand response and other advanced functions, based on the integration of a new two-way communication network, which will be referred as Smart Metering Network (SMN). In this paper, we focus on the design principles of multiple access control (MAC) protocols for SMN. First, we list several AMI applications and its benefits to the current power grid and user experience. Next, we introduces several features of SMN relevant to the design choice of the MAC protocols, including the SMN architecture and candidate communication technologies. After that, we propose some performance evaluation metrics, such as scalability issue, traffic types, delay and etc, and give a survey of the associated research issues for the SMN MAC protocols design. In addition, we also note progress within the new IEEE standardization task group (IEEE 802.11ah TG) currently working to create SMN standards, especially in the MAC protocols aspect.

## 1. Introduction

Traditionally, the collection of power data from end-customer premises has been accomplished by using conventional power meters. Even if such meters could be remotely read as in Automated Meter Reading systems (AMR), such capabilities were limited to one-way, infrequent data upload. The next generation – Advanced Metering Infrastructure (AMI) - is based on a two-way communication network, Smart Metering Network (SMN), which is comprised with the communication transceivers integrated into smart meters. Therefore, in addition to the remote reading function, AMI may also support remote control, load management and demand response by sending down-link control information from the utility. We provide a high-level comparison of these three generations of power metering systems in the Fig.1.

| Generation | System | Comm. Network | Terminal Device | Function |
|---|---|---|---|---|
| Last | Traditional Power Metering System | No Comm. Network | Conventional Meters | Manual Reading |
| Current | Automatic Meter Reading (AMR) | One-way Comm. Network | Conventional Meters Coupled with One-way Comm. Infrastructure | Remote Reading |
| Current/ Next | Advanced Metering Infrastructure (AMI) | Two-way Comm. Network (SMN) | Smart Meters | Remote Reading and Control, Load Management, Demand Response, Fault Detection |

*Figure 1. Comparison among three Generations of Power Metering Systems.*

Since the AMI crucially depends on the two-way networks, the communication aspects of SMN design have begun to draw attentions [5], [23], [24], [7], [17]. Some of those discuss the choice of communication architectures [6] that are appropriate for the various AMI applications, so as to achieve the traditional goals of communication reliability, efficiency and security. On the other hand, some papers talk about the pros and cons of multiple different communication technologies, such as power line communications [25], Zigbee [10] and WiFi [26], and compare their suitability to the AMI applications. Additionally, another key point, which largely determines the efficiency and performance of data communications in SMN is the Multiple Access Control (MAC) protocols, which is exactly the focus of this paper. First, in Section III, we discuss some SMN unique features that will significantly impact the choice of MAC protocols. After that, the MAC performance metrics and their associated research issues are presented in detail in Section IV. The entire paper is concluded in Section V.

## 2. AMI Applications and Benefits

We briefly summarize several representative applications of AMI and their resulting benefits.

### 2.1. Outage Detection and Management

Traditionally, outage may only be detected by a report from customers and other monitoring at the control center. With the aid of SMN, the utility can devise a more efficient way to detect the outage and improve response time. For example, once the voltage drops below a threshold for a duration, the Smart Meter may report a warning message and enable the control center to detect and locate the source of an impending outage event.

### 2.2. Demand Response

With SMN, the customers will be able to acquire their own real-time energy usage data and use dynamic pricing information sent by the utility, to make better decisions or manage automatically on use of the electric appliances within the premises, so as to conserve energy and cost.

### 2.3. Power Quality Monitoring

The status of monitoring of the distribution segment (from the distribution substation to customers) is currently very limited. AMI enables collection of various types of real-time data at many more points - such as feeder voltage and current, power quality within a customer premise etc. This will lead to more efficient and accurate power quality monitoring.

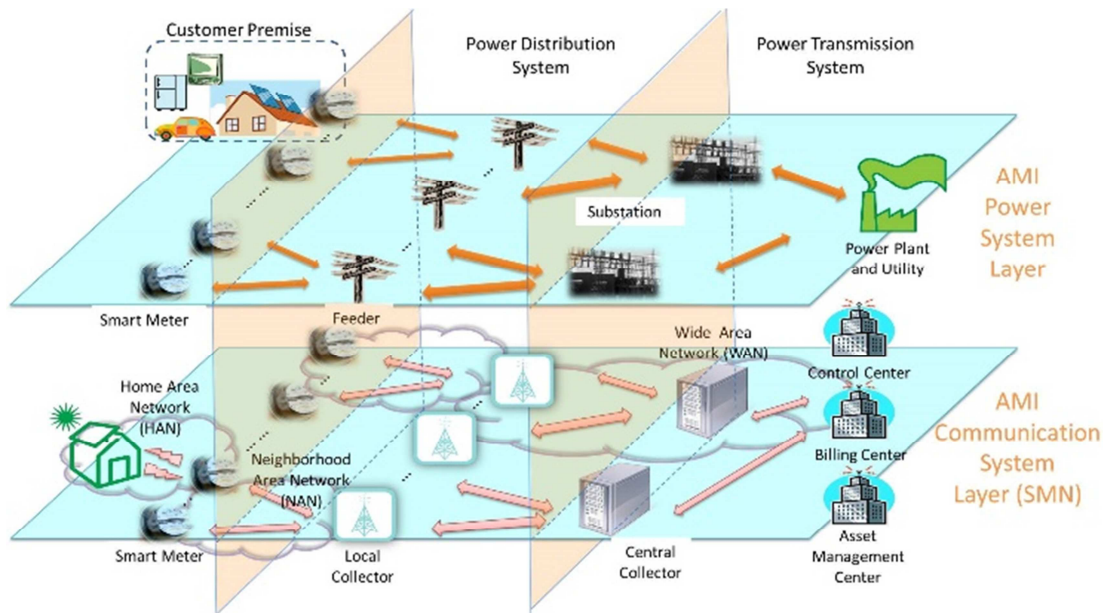

*Figure 2.* AMI Power and Communication System Architectures.

### 2.4. Remote Connect and Disconnect

With the aid of SMN, a utility may remotely enable or disable the energy delivery to certain customers, so as to smooth consumption peaks or automatically react to some emergency events.

The above is only a partial list of applications and benefits from AMI. In summary, with the help of the bi-directional communication network, AMIs can improve operational efficiency, reliability and security.

## 3. Smart Metering Network Features

### 3.1. Smart Metering Network System Architecture

Fig.2 presents an SMN system based on the Smart Grid Architecture given in [1]. A short overview of its main

components follows:

Smart Meter (SM): This device has three different roles. First, the Smart Meter is a multi-utility instrument measuring electric power consumption (and possibly in future, gas, water and heat). It can thus act as an energy control center, i.e. as a point of aggregation for usage information collected by using a Home Area Network (HAN) that connects home appliances. Finally, the Smart Meter also serves as the gateway between HAN and external network; it reports on energy consumption, sends out urgent data, receives remote commands from the utility and is responsible for security of the above transactions. It is noted that SMs nodes are fixed as they are deployed in customer premises. They are usually powered from the main supply and hence power-saving issue is not as important as in traditional battery-powered Wireless Sensor Network nodes.

Home Area Network (HAN): This is composed of multiple inter-connected electric appliances, such as air conditioner, dish washer, plug-in hybrid electric vehicles (PHEV), etc. and the Smart Meter. All the components inside HAN share information or deliver control commands to each other. For example, the dish washer may send a signal to Smart Meter, requesting it to send a `postpone' command to the charging PHEV, so that it may operate without incurring excessive energy cost at that time.

Local Collector: Between the Smart Meter and Utility Center, there could be multiple layers of intelligent electronic devices (IEDs) that acts as data concentrators. For example, a data collection node closer to customer premises - named as Local Collector - collects SM data from multiple premises and relays it to the Central Collector. Additional functions at the Local Collector may include simple data processing and distributed decision and intelligence using its own data. The network segment between Smart Meter and Local Collector is called Neighborhood Area Network (NAN), while that above Local Collector belongs to Wide Area Network (WAN).

Central Collector: A centralized data repository for the entire region operated by the utility that acts as the interface with Control Center, Billing Center and Asset Management Center. These centers may use this data to conduct analysis and evaluate system status, and make decisions or deliver control commands to other components.

The mapping between the SMN components and their prospective physical deployments is also shown in the Fig.2. For example, the Local Collector could be located at a Distribution Transformer because it would be easy to power the Collector and obtain measurements from other feeder devices. On the other hand, the independent deployment for Local Collector may provide more flexibility to adjust its coverage range. The Central Collector is likely to be deployed closer to the centers. If the utility's coverage region is not very large, only one Central Collector located close to the centers may suffice. Otherwise, Central Collectors may be placed at the Distribution Substation as the second tier data relay. Since the HAN consists of electric appliances which are manufactured by different vendors and has much flexibility in implementation especially on application layers, the utility operating SMN may leave it open and focus their design on the upper level network. On the other hand, the design of network from the Local Collector to the Central System does not only depend on the communication requirements of SMN because it also includes the electrical devices which serve the power systems other than AMI. Therefore, the main focus of the MAC protocols design for SMN lies on the segment from Smart Meter to Local Collector (NAN).

### 3.2. Communication Technologies

A large amount of literature discusses the feasibility of several optional bi-directional communication technologies applied on the SMN. We summarize the various options and highlight their pros/cons in Fig.3.

In Power Line Carrier (PLC), the data is transmitted over electricity transmission lines along with electrical power [9]. Its communication performance depends on several factors, such as frequency, propagation distance and existence of transformers because the data signals cannot go through the transformers. PLC has gained a lot of attractions because it uses the existing power lines as signal carrier and no extra cabling fee is needed. Therefore, many countries (e.g. Singapore) adopted it for broadband communication services. However, PLC also suffers from several disadvantages, such as high signal attenuation, high noisy medium and lower scalability, which leads to the termination of deployment in some countries (e.g. US) [8].

ZigBee is a wireless communication technology that consumes low power at the device side [10]. Thanks to its low cost and easy implementation, this technology has already been widely used in the Smart Home network by many AMI vendors, such as Itron and Landis Gyr. They produce smart meters and measuring devices integrated with ZigBee protocol to monitor and control the Home Energy Status. On the other hand, there are still some constraints on ZigBee for its practical application on the SMN. For example, its short range confines this protocol in the application domain of HAN. Furthermore, the processing capabilities and memory size of the ZigBee device are expected to be improved for more advanced functions and communication requirements of the SMN.

Machine Type Communications over Cellular Network allows the Smart Meters and Local Collector to exchange information via low data load communication service, which has been supported by multiple mature cellular network standards, such as LTE [11]. It is the popularity and easy implementation that make this technology become an attractive candidate option. Furthermore, the long range and high data rate provide the utility more flexibility to design and implement the SMN. However, the concern about reliability, security and delay performance makes a barrier for the implementation of this technology in practice, especially under the condition of heavy traffic load.

Finally, WiFi is a communication technology that allows devices to exchange data wirelessly based on IEEE 802.11

Standards. Its popularity, mature development and unlicenced spectrum make it on the top of the candidate technology list. Furthermore, it is also a cost efficient network with dynamic self-healing and distributed control, which makes it easier to be implemented. On the contrary, the capacity, scalability and security issues are the main challenges for its application on the SMN. Therefore, in order to solve these challenges, a new standardization task group IEEE 802.11ah is established and aimed at creating a WiFi based standard to support wireless communication between Smart Meters and Local Collector as one of its primary use cases. According to [14], IEEE 802.11ah compliant devices will utilize Multi-Input Multi-Output, Orthogonal Frequency Division Multiplexing (MIMO-OFDM) at frequencies below 1 GHz, where there is no licensing and regulatory issues. The most discussed channelization for .11ah in the US focusses on the 902 - 928 MHz band, which is currently free. The .11ah Working Group appear to have settled on 1 MHz and 2 MHz as the possible channel bandwidth [13]. In addition to the benefit of free spectrum, the signal transmission below 1 GHz generally suffer less propagation path loss, enabling the network to achieve larger coverage, as verified in the calculations in following section.

| Technology | Applications | Benefits | Limitations |
| --- | --- | --- | --- |
| Power Line Carrier | WAN, NAN, HAN | No Extra Cabling Fee, High Security | High Noisy Medium, Low Scalability |
| Messaging over Cellular Network | HAN, NAN, WAN | Mature Development, Long Range | Low Data Rate, Low Robustness, Low Security, Costly Spectrum Fees, Low Scalability |
| WiFi | HAN, NAN, WAN (with multi-hop) | Mature Development, Free License, High Robustness | Low Security, Low Scalability |
| ZigBee | HAN | Low Cost | Short Range, Low Security, Low Data Rate |

*Figure 3. Comparison among optional Communication Technologies for SMN.*

# 4. MAC Protocols Design Performance Metrics and Important Research Issues

MAC protocols must be designed to match the differing objectives for the various types of Smart Grid data as well as adapt to the different network topology scenarios. Furthermore, the special features and applications in SMN also address some new challenges to the suitable MAC protocols design. We next outline the broad performance metrics for MAC protocols as they relate to Smart Grid operations, and identify some specific research challenges for SMN MAC protocols.

## 4.1. Different Data Types

Usually, the Smart Meter traffic may be classified into two different classes: periodical and event-triggered data. The former includes energy consumption information while the latter is largely data from protection devices (relays, reclosers etc. that monitor local fault status) or electric vehicle charging stations. We list a table (Fig.4) of several representative traffic examples with their important properties. Given the different requirements and features of traffic types, different MAC protocols should be designed specially for each traffic type. For example, [28] targets on the event-driven data, while [27] is designed for the periodical reporting data.

| Traffic Examples | Delay Requirements | Trigger Type |
| --- | --- | --- |
| Outage Alert | Seconds | Event Triggered |
| Billing Information | Minutes to Hours | Periodical |
| Demand Response | Seconds or Minutes | Periodical |
| Real-time Pricing Information | Seconds or Minutes | Periodical |
| EV Charging Information | Seconds or Minutes | Event Triggered |

*Figure 4. Examples of Data with Different Communication Requirements.*

## 4.2. Delay

Different types of data induce different communication requirements within a SMN. For example, the energy consumption data is delay tolerant. On the other hand, the delay sensitive data - such as those reporting a fault and protection related messages - should have higher priority over others, so as to minimize end-to-end latency. Therefore, how to minimize the SMN latency - such as that of the last hop between the Smart Meter and the Local Collector is a primary concern. In general, several factors impact the delay, such as the choice of the communication technology, the network architecture, and most notably, the MAC protocol. A good MAC protocol can coordinate the uplink transmissions of multiple communication nodes to reduce the collision probability significantly, resulting in lower delay.

Usually, it is convenient to design MAC protocols for one type of traffic. For example, a polling based (taking-turns) protocol such as Point Coordination Function (PCF) defined in 802.11 is well-suited to reporting data with bounded delay guarantees. The total duration of one polling cycle increases only linearly with the number of nodes (in contrast to exponential increase in delay with random access systems as the aggregate load increases) and provides a guaranteed delay bounds. However, the efficiency of such protocols declines rapidly with the number of nodes. On the other hand, random access MAC protocols such as Distributed Coordination Function (DCF) in 802.11 have been designed to provide reasonable efficiencies in terms of throughput at low to moderate loads, but the delays escalate rapidly as the average load increases. In order to solve this challenge, [28] proposes two grouping based DCF MAC protocols: TDMA-DCF and Group Leader DCF-TDMA scheme. These two schemes divide all Smart Meters into several groups to reduce the competitive channel access and are both directed at 802.11 type networks operating at the frequencies below 1GHz, which has been adopted by the IEEE 802.11ah TG.

However, most Smart Grid scenarios comprise of a mix of traffic, e.g. regular traffic and emergency traffic. To serve both types within a DCF framework, the notion of traffic classes were introduced via Enhanced Distribution Channel Access (EDCA) defined in 802.11e [4], to prioritize low-latency (event-driven) data over non time-critical data applications (such as email). A combination of EDCA and PCF, Hybrid Coordination Function Controlled Channel Access (HCCA) tries to serve multiple traffic types by granting higher priority to some particular kinds of data via polling algorithm, which is centralized controlled by Access Point (AP). On the other hand, the performance against scalability issue of these hybrid MAC protocols under densely populated network still needs to be evaluated.

*4.3. Scalability*

Scalability, which is the most significant challenge in SMN, requires that MAC protocols continue to perform well as the number of SMs offering data scales. Referring to [16], one Local Collector is required to support a network with up to 6000 SMs when considering MAC performance. According to [15], the urban outdoor path loss model for 900MHz RF in dB is $P_{L,dB}(r) = 8 + 37.6 \times \log_{10} r$, where $r$ denotes the distance between the Smart Meter and Local Collector. Then based on the parameters listed in the table (Fig.5) [14], the received power and noise power at the receiver terminal are given as:

$$P_{RX,dB}(r) = P_{TX,dB} + G_{dB} - P_{L,dB}(r)$$
$$= 0 + 3 - (8 + 37.6 \times \log_{10} r) \quad (1)$$

$$P_N = k \times T \times W = 1.3 \times 10^{-23} J/K \times 290K \times 2MHz \quad (2)$$

| Parameter | Value | Parameter | Value |
|---|---|---|---|
| Channel Bandwidth $W$ | 2 MHz | Transmitter Power $P_{TX,dB}$ | 30 dBm |
| Antenna Gain $G_{dB}$ | 3 dB | Temperature $T$ | 290 K |
| Transmission Frequency | 900 MHz | Boltzmann's Constant $k$ | $1.38 \times 10^{-23}$ J/K |
| SM Density $\rho$ | 1000 per $km^2$ | Cell Radius $R$ | 1200 m |
| $SINR_{dB,th}$ | 14 dB | $P_{RX,dB,th}$ | -129 dB |

*Figure 5. Parameter List for Transmission Rate and Hidden Node Calculation.*

Referring to the AWGN capacity derivation with BPSK modulation in [18], we may draw a figure of the transmission rate and received power with respect to $r$. As shown in the Fig.6, the RX received power at 1200m satisfies a reasonable received power threshold (-120dB) and achievable transmission rate at 4500m exceeds 100kbps, the minimal required data rate set in IEEE 802.11ah Use Case [16]. This implies that one Local Collector may need to communicate with individual Smart Meters located 1200m away using a star topology. Given typical SM density $\rho$ (1000-6800 SMs per km²) [21], [22], it is possible to have in excess of 6000 Smart Meters communicating to one Local Collector. Clearly, the design of MAC protocols for such large number of nodes, invites challenges about scalability of any chosen MAC protocol for SMN as discussed next.

For random access protocols based on carrier sensing - such as DCF (CSMA/CA) in IEEE 802.11 - such a large coverage area corresponding to a single Collector cell will lead to significant hidden nodes. In the traditional random access protocols, such as CSMA/CA, all nodes listen to estimate channel status (busy/idle) based on energy thresholding. If the node observes the channel to be continuously idle for a specific interval, it starts contending for the channel via a random back-off process. However, when the coverage of the Local Collector is enlarged, one Smart Meter (hidden node) may not be able to detect ongoing transmission of other meters due to the degenerated radio channel condition. Then this hidden node may initiate its own transmission because it determines the channel as idle, which leads to a collision.

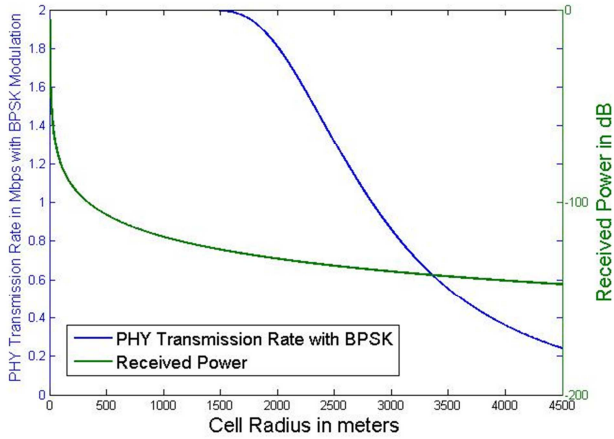

*Figure 6. PHY Transmission Rate of the Communication with BPSK Modulation and Received Power at Local Collector with respect to Their Distance.*

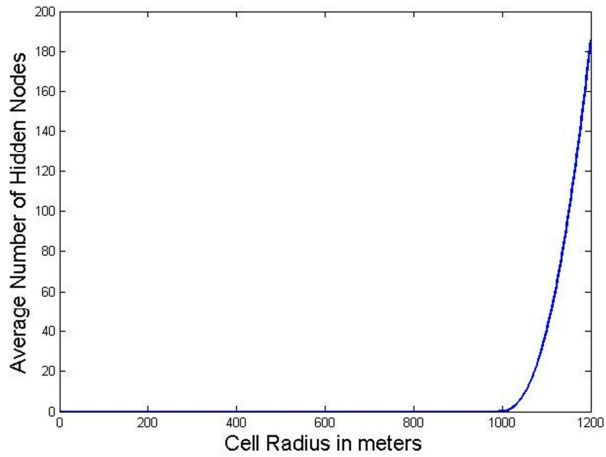

*Figure 7. Network Topology for Hidden Node Calculation.*

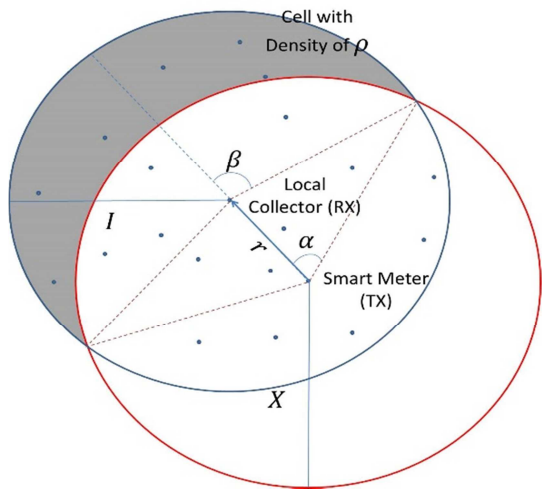

*Figure 8. Results for Hidden Node Calculation.*

In order to investigate the number of hidden nodes with growing Local Collector coverage, we consider the network topology of a disk with radius of $R$ and uniformly distributed SM density $\rho$ (Fig.7), the distribution of Smart Meter deployment with respect to $r$ is $f(r) = 2r/R^2$, where $0 \leq r \leq R$. Since the down-link communication from Local Collector may cover all the Smart Meters successfully, thus the hidden node only exits in up-link communications where Smart Meter is always the transmitter (TX) and Local Collector is always the receiver (RX). According to [20], the area inside the interference range of the receiver $I$ and outside the carrier sensing range of the transmitter $X$ is defined as the hidden area $A(r)$ (shadowed zone) for a given Smart Meter, which is given as:

$$If\ r \leq X - I, A(r) = 0$$

$$If\ r > X - I, A(r) = \beta I^2 + rX|\sin(\alpha)| - \alpha X^2 \quad (3)$$

where $\alpha = \cos^{-1}(X^2 + r^2 - I^2/2rX)$ and $\beta = \pi - \cos^{-1}(I^2 + r^2 - X^2/2rI)$.

The Carrier Sensing Range $X$ can be calculated based on the equation $P_{TX,dB} + G_{dB} - P_{L,dB}(X) = P_{X,dB,th}$, where $P_{X,dB,th}$ is the carrier sensing threshold for the received power such that the received signal can be detected. Furthermore, the Interference Range $I$ can be derived based on the following equation:

$$I(r) = \min[R, \{y | P_{RX,dB}(r) - 10\log_{10}(P_N + P_{RX}(y)) = SINR_{dB,th}\}] \quad (4)$$

where $SINR_{dB,th}$ is the threshold for Signal Noise Interference ratio such that received signal can be decoded successfully.

After that, we use $N_{hidden} = \int_0^R \rho A(r) f(r) dr$ to obtain the mean number of hidden nodes inside the network and plot it with respect to cell radius $R$ based on the parameters listed in Fig.5. As shown in the Fig.8, the number of hidden nodes increases dramatically when the coverage of the network is enlarged. Although the DCF in 802.11 proposes the RTS/CTS algorithm to reduce the occurrence probability of hidden nodes event, its effect on such a large area network still needs to be investigated. On the other hand, the increasing number of Smart Meters may also increase the data load in the SMN, which basically results in stronger competition for the medium access. Then the collisions and following retransmissions happen more frequently, which directly aggravate the performance.

In order to solve this scalability issue, the IEEE 802.11ah TG is considering using improved DCF and PCF [14]. The modified DCF with Contention Factor and Prohibition Time is one of the suggestion new options. Before the contention phase, the Local Collector broadcasts a prohibition time $T$ and contention factor $0 < Q < 1$, according to the current network congestion status. After that, each Smart Meter generates a random number $r$ which follows a uniform distribution on the unit interval and compare it to $Q$. Then the Smart Meter may contend for the channel if $r \leq Q$ and, otherwise, it keeps silent until the prohibition time $T$ passes. In order to further relieve contention congestion, the MAC scheme may also divide all the Smart Meters within a cell into several groups and provide different groups with different parameters $Q$ and $T$ or allow them to contend for the channel group by group. The collision probability is expected to decrease dramatically (compared to traditional DCF applied to all SMs), as a result.

In order to solve the scalability challenge for PCF, IEEE 802.11ah TG proposes a modified PCF scheme, Probe and Pull MAC (PP-MAC) [14]. After partitioning all the SMs into groups, the Local Collector broadcasts a Probe message to a certain group of Smart Meters before the contention free phase. After that, the Smart Meters having data to send reply a short Probe-ACK concurrently with the use of Zadoff-Chu sequences. By assigning these orthogonal sequences to each Smart Meter and multiplying their own messages with their respective sequences, the cross-correlation of the simultaneous short Probe-ACK transmissions is reduced, so that the Local Collector is able to resolve these parallel ACKs and identify the different transmitters. After that, the Collector schedules and only polls the Smart Meters with Probe-ACK, which leads to a shorter polling cycle and a more efficient taking-turns MAC protocol.

### 4.4. Fairness

This seeks to measure whether each node in the SMN obtains a fair share of system resources. Fairness can be quantified in terms of the access probability to the shared channel by each node - ideally, this should be equal (independent of the node) assuming that all SMs require identical data rates. In general, the notion of Proportional Fairness should be applied, based on different data rate requirements by different nodes. For example, the Least Completed First Send (LCFS) Principle proposed in [27], provide an efficient way to guarantee the fairness of access opportunity of each SM.

### 4.5. Security

Data security in SMN is an extremely vital issue as it relates to household or customer information (e.g. energy consumption profile) that is considered private. Therefore, it is necessary to encrypt the message to prevent eavesdroppers from intercepting the message. Although cryptographic tools and algorithms are relatively mature, these will result in extra load on the SMN. An open question is whether the known features of SM data may be exploited to develop simplified yet effective cryptographic approaches. Secondly, end-point authentication is also indispensable for SMN; whenever the data collector receives a message of energy consumption report, it has to authenticate the identity of the sender. Specifically, defenses against two common types of attacks will be of high priority. Integrity in data communications between SM and Local Collector may be compromised by a relay or man-in-middle attacker. And such communications may be targeted for disruption via Denial-of-Service (DoS) attacks by saturating the Local Collector with a large number of spurious external communications, so that it cannot respond to the legitimate traffic [12]. Within this context, it is noted that most SM communications are regular as it reporting actions are typically scheduled. Therefore, we may exploit such features to filter out malicious accesses by an attacker, by identifying anomalous access traffic patterns.

### 4.6. Expandability

For SMN, the expandability means the ability of this network to accommodate the new communication nodes (Smart Meters) to its existing capacity. It is noted that the deployment of Smart Meters will not occur according to a fixed schedule; for example, whenever a house is built, a newly installed Smart Meter will need to be introduced into the existing NAN covered by the corresponding Local Collector. This introduction procedure, which may include registration, identity authentication, geographical location identification, has to be conducted automatically and is used for the Local Collector to determine the newly installed Smart Meter is ready to work inside its coverage. Therefore, how to realize this introduction procedure should be a part of the MAC protocols design. For example, whenever one newly installed Smart Meter is online, it sends a request to its associated Local Collector to report its own identification and geographical location. After that, the collector registers this new meter and replies it via a message with some necessary setting information. Then how to automatically modify the parameters of current communication systems due to the newly registered Smart Meter still needs to be analyzed.

### 4.7. Fault Detection

For SMN, the fault can be categorized into two kinds of cases. The first one is data fault, which means the data involved in the message has some errors. These data errors may be caused by monitoring errors or malicious message altering. Fortunately, the data collector may detect such kind of fault by some statistical algorithms, such as comparison between the current data and historical data. This kind of fault and corresponding solutions mainly occur at application layer. What the MAC protocol designers need to consider is another one, communication fault, which means that some Smart Meters cannot communicate with the Local Collector directly. These communication problems may be caused by the malfunction of Smart Meters or the degeneration of wireless communication environment. Since communication fault leads to the loss of message, which is intolerable for SMN, it is required to design a MAC protocol to detect the `silent node' inside the SMN as quickly as possible. For example, the Local Collector may exploit the idle communication intervals to poll every Smart Meter and expect its feedback. After that, the collector may detect the silent meters by checking the missing feedbacks. Furthermore, how to schedule the poll-feedback actions in detail and improve its efficiency still need to be investigated.

## 5. Conclusion

This paper highlighted the design consideration for the SMN, the two-way communication network for next generation AMI systems. We listed several important AMI applications and their benefits to the current power grid. The features and requirements of SMN which have much impact on its MAC protocols design were discussed, which includes the SMN architecture and some candidate communication technologies. Most importantly, we highlighted several important MAC performance evaluation metrics, including traffic types, delay, scalability, fairness, security and expandability, and also gave a survey of associated research issues for SMN MAC protocols, including a summary of current status for MAC protocol design effort in IEEE 802.11ah TG.